\documentclass[prl,twocolumn,groupedaddress,showpacs,nofootinbib]{revtex4}
\usepackage{amsmath}%
\usepackage{bm}%
\usepackage{latexsym}%

\newcommand{\lpl}{$l_{\rm pl}~$}
\newcommand{\mpl}{m_{\rm pl}}

\begin{document}
\bibliographystyle{apsrev}

\title{Trans-Planckian dispersion and scale-invariance of
inflationary perturbations}

\author{Jens C. Niemeyer}
\email[]{jcn@mpa-garching.mpg.de}
\affiliation{Max-Planck-Institut f\"ur Astrophysik,
Karl-Schwarzschild-Str. 1, D-85748 Garching, Germany}
\author{Renaud Parentani}
\email[]{parenta@celfi.phys.univ-tours.fr}
\affiliation{Laboratoire de Math\'{e}matiques et Physique
Th\'{e}orique, CNRS UMR 6083,
Universit\'{e} de Tours, 37200 Tours, France}


\begin{abstract}
We investigate the insensitivity of the  predictions of
inflationary models with respect to modifications of Planck energy
physics. The modification we consider consists in replacing the usual
dispersion relation by nonlinear ones. This way of addressing the
problem has recently recieved attention and contradictory results
were found. Our main result is to show that the adiabaticity of the
mode propagation  and the separation of two scales of interest, the
Planck scale and the cosmological horizon scale, are sufficient
conditions for the  predictions to be unchanged. We then show that
almost all models satisfy  the first condition if the second is met.
Therefore the introduction of a nonlinear dispersion  is unlikely to
have any discernable effects on the power spectrum of cosmological
perturbations. 

\end{abstract}


\maketitle

A number of recent publications \cite{J00,MB00,N00,KG00,K00} address
the so-called trans-Planckian problem of inflationary cosmology:
in some inflationary scenarios, scales of cosmological interest today
were redshifted from wavelengths far below the Planck length, \lpl,
at the onset of inflation.  Since we do not have at present a
consistent quantum theory  at these scales, this appeal to unknown
physics questions the validity of the standard predicitions of
inflation,  i.e.~scale invariance of the power spectrum of cosmic
fluctuations and Gaussianity of their distribution.

In usual scenarios, the metric fluctuations
responsible for the generation of large scale structure and the
temperature anisotropies of the cosmic microwave background radiation
(CMBR) originate from quantum fluctuations of a massless field
propagating in a
background geometry whose expansion rate $H$ is determined by the
potential energy of the inflaton field \cite{L84,KT90}.  More
specifically, the power spectrum of the metric fluctuations  can be
obtained from the vacuum expectation value of the two-point function
of a minimaly coupled scalar field,  evaluated at the time when the
field wavelength is equal to the cosmological horizon $H^{-1}$
\cite{M88}.  If the potential is sufficiently flat, $H$ is nearly
constant in time, giving rise to a nearly scale invariant
horizon crossing amplitude.  Moreover, the linearity of the field
equation results in the Gaussianity of the distribution statistics
when the state of the field is vacuum.

To question the robustness of these results with respect to changes of
Planck energy physics, Refs.~\cite{MB00,N00,KG00} modified the
propagation of the  quantum field in this high frequency domain,  in
close analogy to what was done for Hawking radiation (see \cite{J00}
for a review and references).  The modification consists in replacing
the  linear dispersion relation which follows from Lorentz
invariance by ad hoc nonlinear dispersion relations. The linearity of
the field equation was maintained, guaranteeing that
Gaussianity be preserved.  Different forms of the dispersion relation
were investigated,  all of which reducing to the linear behavior for
small frequencies but becoming nonlinear above a given  wavenumber
$\kappa_c$  which can be thought to be of the order of the inverse
Planck length.  Assuming that horizon crossing occurs in the linear
regime of the dispersion relation (i.e., that $H \ll \kappa_c$),
Refs.~\cite{N00,KG00} find no violation of scale invariance of the
fluctuation power spectrum. On the other hand, Ref.~\cite{MB00} claims
to find significant scale non-invariance in some cases.

These opposite results raise three questions we shall address
in this Letter: 
(1) What are the {sufficient} requirements  for a
nonlinear dispersion relation {\sl not} to significantly violate 
 the standard predictions of inflation? 
(2) How do the first deviations from the
standard results manifest themselves in the power spectrum ?
(3) Is the cosmological situation
different from the black hole case \cite{U95,BMPe95,CJ96,SS00} 
where no deviation in the thermal
spectrum was found when the cutoff frequency is much higher
than the Hawking temperature, and if yes, why?

Given the spatial homogeneity of the background geometry, the
propagation of a linear field is governed by  the time evolution of
the field modes $\chi_k$ with conserved comoving wavenumber $k$. 
Therefore the $k$-th component of the power spectrum is
determined only by the amplitude of $\chi_k$ at the (conformal) time
$\eta_k$ of horizon crossing, i.e., when $k = a(\eta_k)\,H$, where
$a(\eta)$ denotes the scale factor of the Friedmann-Robertson-Walker
metric.  When using the conventional dispersion relation, $\chi_k$
obeys the equation
\begin{equation}
\label{mode}
\chi_k'' + \omega^2 \chi_k = 0\,\,,
\end{equation}
with $\omega^2$ given by
\begin{equation}
\label{omega}
\omega_0^2(\eta) =  k^2 - \frac{a''(\eta)}{a(\eta)} \, ,
\end{equation}
where dashed variables stand for derivatives with respect to $\eta$
(hence, $H=a'/a^2$).  Upon modifying the dispersion relation, the wave
equation  is still given by Eq.~(\ref{mode}) but $\omega_0^2$  is
replaced by 
\begin{equation}
\label{omegaF}
\omega_F^2(\eta) = [a(\eta) \,F(k/a(\eta))]^2 -
\frac{a''(\eta)}{a(\eta)} \, .
\end{equation}
$F(k/a)$ is an as yet unspecified function which converges to the
usual  physical wavenumber $\kappa= k/a$ for $\kappa \ll \kappa_c$. 

The phase and amplitude of $\chi_k$ are fixed by two conditions.
First, its normalisation is provided by the Wronskian
condition,
\begin{equation}
\label{wronski}
\chi_k \, {\chi_k'}^\ast - \chi_k^\ast \, \chi_k' = i \,\,,
\end{equation}
which follows from the requirement that the creation and annihilation
operators satisfy the canonical commutation relations \cite{BD84}.
This condition equally applies for $\omega_0$ and $\omega_F$.
Secondly, one needs a positive frequency condition.
Then the mode so defined uniquely specifies the creation operator
$a^\dagger_k$  that can be used to construct 
the $k$-component of the field state in Fock space.

The physical requirement fixing the state 
at $\eta_i$, i.e., at the onset of inflation, 
is that the $k$-components corresponding 
to physical wavenumbers $\kappa_i=k/a(\eta_i)$ 
that are very large, i.e., well above the temperature and $H$,
should be in their ground state.
When using the usual frequency $\omega_0$ of Eq. (\ref{omega}),
there is no problem in charcaterizing the
vacuum since  $\omega_0$ is constant for early $\eta_i$.
However, $\omega_F $ might still be time dependent at $\eta_i$.
Hence one looses the conventional way of defining the vacuum.
In order to define it,
we shall follow the conventional adiabatic approach \cite{BD84}. 
In an adiabatic regime,  the adiabatic vacuum equivalently follows 
from the mathematical properties of the modes $\chi_k$ \cite{MP98}
or from the physical requirements to minimize either 
the probability for excitation of a comoving particle detector 
or the energy density of the field \cite{MB00}. This is
because both quantities are based on the local properties 
of $\chi_k$ that are accurately described by WKB waves.

It is well known that the adiabatic vacuum is associated with 
the positive frequency WKB solution of
Eq.~(\ref{mode}), normalized according to Eq.~(\ref{wronski}) and given by  
\begin{equation}
\label{WKB}
\chi^{\!W\!K\!B\!}_{k}(\eta) = \frac{1}{\sqrt{2 \omega(\eta)}}
 \exp\left(-i  \int^\eta_{\eta_i} \omega(\tilde \eta) d\tilde
 \eta\right)\,\,.
\end{equation}
The key point for us is that the probability for {\it non-adiabatic}
transitions \cite{MP98} (i.e., transitions leading to spontaneous
excitation of the vacuum and interpreted in second quantization as
pair creation) is determined by the corrections to Eq. (\ref{WKB}).
Indeed, in the adiabatic regime, i.e.~when the adiabatic  parameter
\begin{equation}
\label{condition}
{\cal C}(\eta) \equiv \left|\frac{\omega'}{\omega^2}\right| 
\end{equation}
satisfies ${\cal C}(\eta) \ll 1$ for all $\eta_i < \eta < \eta_f$, 
the pair creation probability amplitude at $\eta_f$ is equal to 
backscattering amplitude $\beta_k$, i.e.~the amplitude
of the negative frequency part of $\chi_k$ evaluated at $\eta_f$,
when starting at $\eta_i$  with the positive frequency mode
Eq.~(\ref{WKB}). Explicitely, $\beta_k$ is defined by decomposing 
the exact solution $\chi_k$ at $\eta_f$ as
\begin{equation}
\label{decomposition}
\chi_k(\eta_f)= \alpha_k \chi^{\!W\!K\!B\!}_{k}(\eta_f)
+ \beta_k [\chi^{\!W\!K\!B\!}_{k}(\eta_f)]^* \, ,
\end{equation}
where $ \vert\alpha_k \vert^2 - \vert \beta_k\vert^2 =1$. 

In the context of inflation, if the evolution
from $\eta_f$ to the horizon crossing time $\eta_k$
is governed by $\omega_0$ (see below) 
$\beta_k$ modifies the power spectrum
$(\Delta \Phi)_k^2 \sim k^3 |\chi_k(\eta_k)|^2/a^2$.
Using Eq.~(\ref{decomposition}), we find
\begin{equation}
\label{modulation}
\vert \chi_k(\eta_k) \vert^2 =\vert   \chi^{(0)}_{k}(\eta_k)
\vert^2 \, \,  [ 1 + 2 \Re e ( \beta_k e^{i \phi_k} )
 + O(\beta^2)] 
\end{equation}
where $ \chi^{(0)}_{k}$ is the unperturbed amplitude evaluated 
using $\omega_0$ when working in the unperturbed 
vacuum ($\beta_k =0$). $e^{i\phi_k}$ is a pure phase which depends on the
value of $\eta_f$. Hence $\beta_k$ determines the amplitude of the
modulation around the usual horizon crossing amplitude. 

The condition for Eq. (\ref{modulation}) to be valid is that 
the horizon and cutoff scales be well separated:
\begin{equation}
\label{sigdef}
\sigma \equiv \frac{\kappa_c}{H} \approx \frac{\mpl}{H} \gg 1\,\,.
\end{equation}
In this case, $\eta_f$ can be chosen such that the physical
wavenumber $\kappa=k/a$ is,  at the same time, much smaller than
$\kappa_c$ and much bigger than $H$. In this intermediate regime, the
second term in Eqs.~(\ref{omega}) and (\ref{omegaF}) is small compared
to the first for $\eta < \eta_f$ and can therefore be neglected in the
evaluation of ${\cal C}$. Furthermore, $F$ is assumed to be in the linear
regime at $\eta_f$, thus $F(k/a(\eta > \eta_f)) = k/a$. The
evolution of $\chi_k$ from $\eta_f$ to  $\eta_k$ is therefore
identical to the standard, non-dispersive case.
Consequently, to obtain identical horizon crossing amplitude, 
it {suffices} to show that the modified mode 
governed by $\omega_F$ coincides at $\eta_f$ with
the purely positive frequency mode obtained when using $\omega_0$. 
This requires only $\beta_k \ll 1$.

In the case where, in addition to ${\cal C} \ll 1$, ${\cal C} \to 0$ 
for both $\eta \to \eta_i$ and $\eta \to \eta_f$, $\beta_k$ 
is exponentially small \cite{MP98}. 
In our specific case governed by $\omega_F$ this exponential would be
of the type  $e^{-\kappa_c/H}$, i.e.~completely negligible if
Eq.~(\ref{sigdef}) is satisfied. However, the condition 
${\cal C} \to 0$ for $\eta \to \eta_f$ is unlikely to 
be met since one has to evaluate $\beta_k$
at a given finite time, in particular before the last term of
Eq.~(\ref{omegaF}) becomes significant. 
When evaluating $\beta_k$ with ${\cal C}(\eta_f) \neq 0$, only
weaker predictions can be made. 
Nevertheless, when ${\cal C} \ll 1$ and slowly
varying  from $\eta_i$ till $\eta_f$, $\beta_k$ is bounded
by the maximum of ${\cal C}(\eta)$. (This can be 
obtained from the boundary contributions at $\eta_f$ 
generated by multiple integrations by parts, see
Sect. 2.4 in \cite{MP98}.) 
Hence $\beta_k$ is still negligible when ${\cal C(\eta)} \ll 1$. 

In brief, the robustness of the  preditions of
inflationary models follows from (a) the linear relationship 
between the modulations of the horizon crossing amplitude
and the backscattering amplitude and (b) from the smallness of this
amplitude when  both adiabaticity, ${\cal C} \ll 1$, and 
scale separation, $\sigma \gg 1$, are met. 

This is in complete agreement with what has been found in  black hole
physics. In that case, due to the stationarity of the  metric, one
works with fixed frequency $\omega$ rather than fixed wavenumber $k$
and the differential analysis proceeds with the radial variable $r$
rather than with the time $\eta$.  When the appropriate adiabatic
condition is met, see Eq. (47) in \cite{BMPe95}, and when the cutoff
scale is well separated from the Hawking temperature,  which is fixed
by the inverse radius of the hole and which plays the role of $H$,
positive norm WKB waves  provide good approximations of the modified
modes,  in the sense that the amplitude of  negative  norm component
of these modes is negligible.  This guarantees that the modifications
of the Hawking flux will be equally negligible.  In fact, the first
order modulations of the  flux have been shown to scale like
$\sigma^{-2}$ \cite{SS00}.  Thus, as in cosmology, scale separation
and adiabaticity of the mode propagation  protect the predictions from
alterations generated by a nonlinear dispersion relation.

Let us now show that in most cases, scale separation Eq. (\ref{sigdef})
leads to adiabaticity through the relation ${\cal C} \simeq
\sigma^{-1} \ll 1$. To this end, we write Eq.~(\ref{condition}) as 
\begin{equation}
\label{cond3}
{\cal C} \approx \left|\frac{H}{F} - \frac{H \kappa}{F^2}
\frac{d\,F}{d\,\kappa}\right| \le \left|\frac{H}{F}\right| +
\left|\frac{H \kappa}{F^2} \frac{d\,F}{d\,\kappa}\right|
\end{equation}
with $\kappa=k/a$.  Obviously, the  dispersion relation used by
Unruh \cite{U95} which behaves as $F(\kappa) \to \kappa_c$ for $\kappa \to
\infty$ is globally adiabatic as the first term is $\sim \sigma^{-1}
\ll 1$ and the second term vanishes. This is in agreement with the
results of Refs.~\cite{MB00,N00}. 

We consider next the class of functions $F(\kappa)$  which grow
monotonically with $\kappa$.  Here, the first term in
Eq.~(\ref{cond3}) can be neglected as it is smaller than the
corresponding term obtained using the linear dispersion relation and
which is already  negligible. The second term also  vanishes
asymptotically for $F \sim \kappa^n$ with any positive $n$,
as well as for exponentially and logarithmically growing
functions. This class includes the dispersion relation of
Ref.~\cite{KG00} in agreement with the author's conclusion.  On the
other hand, it also includes  the superluminal dispersion relation
investigated in Ref.~\cite{MB00} where it is claimed that it gives
rise to deviations from the standard result.  Our result is obviously
in conflict with that of Ref.~\cite{MB00}. 

Finally, the dispersion relation may become singular at a finite
wavenumber $\kappa_\infty \simeq \kappa_c$, 
\begin{equation}
\label{singular}
F(\kappa) = \frac{\kappa}{(1 - \kappa/\kappa_\infty)^n}\,\,.
\end{equation}
For $n = 1$, the second term in Eq.~(\ref{cond3}) becomes equal
to the first  at $\kappa = \kappa_\infty$, whereas it vanishes for
$n > 1$ and diverges  for $n < 1$. Only in the latter case,
the notion of an adiabatic vacuum becomes invalid near the cutoff
scale, making the predictions for inflationary fluctuations dependent
on an alternative choice of initial conditions, or a later
specification of the initial state, see below. 

The situation is different if $F$ decreases monotonically for 
$\kappa > \kappa_0 \simeq \kappa_c$, 
as the first term in Eq.~(\ref{cond3}) grows without
bound and ${\cal C} \ne 0$ for nonlinear $F$. Even for strictly
positive $F$, i.e. disregarding the additional complications due to
complex dispersion relations, there exists a time $\eta$ where $F$ is
smaller than $H$. According to Eq.~(\ref{cond3}), this causes the
breakdown of the adiabatic condition because cosmological expansion
begins to dominate the dynamics. In other words, those modes with
wavenumber $\gg H^{-1}$ are ``frozen'' in the beginning \footnote{This
phenomenon was also investigated in Ref.~\cite{MBP01} which appeared
shortly before this article was finished.}.  As they are
streched by cosmic expansion,  their proper frequency grows according
to the nonlinear behavior of the dispersion relation.  Therefore,
because of Eq. (\ref{sigdef}), the adiabaticity condition will be
satisfied before the dispersion relation reaches the linear branch.
As an example, consider the subluminal dispersion relation  of Corley
\& Jacobson \cite{CJ96} (also analyzed in Ref.~\cite{MB00}):
\begin{equation}
F(\kappa) = \kappa \sqrt{1 - \left(\frac{\kappa}{\kappa_0}\right)^{2
n}}\,\,.
\end{equation} 
At $\kappa = \kappa_0/2$,  the corresponding adiabaticity parameter is
${\cal C} \simeq   \sigma^{-1}  \ll 1 $ for $n=O(1)$.

If an early non-adiabatic phase is followed by an intermediate
adiabatic regime on Planckian but sub-horizon scales, any viable
initial solution cannot deviate too far from the adiabatic vacuum
without violating the consistency of the inflationary framework.  In
the intermediate adiabatic regime, the number of  these primordial
particles is constant and the usual rules for computing the energy
density apply.  Therefore their energy density will be of the order of
$\kappa_0^4  n(k)$ where $n(k) =\vert \beta_k \vert^2$ is their
occupation number density.  Without some fine tuning of $n(k)$, this
term may become  comparable to or even larger than the inflaton
potential energy,  invalidating the key condition for inflation to
take place. This point was recently (and independently) developed by
Tanaka \cite{T00}\footnote{For an early discussion of this point in the
black hole context we refer to \cite{J91}. In this case, for exactly
the same reasons, the hypothesis of having a regular 
metric near the horizon severly constrains the occupation
number $n(\omega)$ if an intermediate adiabatic regime exists.}.  On
the other hand, if the contribution of $n(k)$  to the expansion rate
is negligible compared to that of the inflaton potential, one has $n(k) \ll
\sigma^{-2} \ll 1$ for $k/a \simeq \kappa_c$,
i.e., these modes must be extremely close to the adiabatic vacuum.
Thus the predictions of inflation remain unchanged.  These remarks
apply, for instance, to the case $n < 1$ in Eq.~(\ref{singular}) if
$\kappa_\infty/H \ll 1$ and to subluminal dispersion relations with $F
\le H$. 

To summarize, we found that all regular, monotonically growing
dispersion relations are globally adiabatic and thus unlikely to give
rise to non-standard perturbation spectra if horizon crossing occurs
in the linear dispersion regime.  Only those functions that decrease
with increasing wavenumber, or that contain a certain type of
singularity, violate adiabaticity at early times. In those cases, the
specification of the initial state is conceptually less clear than in
the former where the adiabatic vacuum is available. However, if
adiabaticity is restored well before horizon crossing, the initial
conditions can be assigned in this intermediate regime, guaranteeing
that the adiabatic solution holds after this point.  Again, the
standard inflationary predictions remain  unchanged. 

One can also question the validity of the assumption of scale
separation itself, and speculate about potential signatures if the
field is still in the (slightly) nonlinear dipersion regime at horizon
crossing. Given that in the simplest scenarios for inflation, the CMBR
fluctuation amplitude scales inversely with $\sigma$ evaluated at
horizon crossing of the respective wavelength \cite{L84}, we know that
$\sigma \approx 10^5$. Whether or not this number should be considered
``large'' cannot be answered without independent experimental
constraints on residual Planck scale effects on long wavelength
phenomena. Indeed, Amelino-Camelia \& Piran \cite{AP00} invoke
nonlinear corrections to the dispersion relation as a possible
explanation for the observation of ultra-high energy cosmic rays and
TeV-photons whose fluxes would normally be suppressed by interactions
with CMBR and infrared background photons,
respectively. Intriguingly, their allowed region of parameter space is
consistent with a second order correction whose coefficient
corresponds to $\sigma \approx 10^5$. Nonlinear dispersion acting on
length scales comparable to the horizon length during inflation
therefore cannot be ruled out on experimental grounds alone. By noting
that the amplitude of the WKB solution Eq.~(\ref{WKB}) depends only on
the local value of $F$ rather than its history, one can estimate the
ratio of the horizon crossing amplitude in the standard and dispersive
theories as $\sim H/F(H)$ \cite{N00} (assuming strict adiabaticity of
the prior evolution). Like in the standard theory, time  translation
symmetry of de Sitter space has to be broken by making $H$ a weakly
time dependent variable in order to obtain a scale dependent
perturbation spectrum. Depending on $g = \Delta F(\kappa)/\Delta
\kappa$, with $\Delta k = \Delta \kappa \,a$ covering the cosmological
scales observable today, we speculate that the natural reddening of
the spectrum due to the slow decline of $H$ will be slightly enhanced
($g > 1$), reduced ($0 \le g < 1$), or even reversed ($g < 0$) as the
fluctuation amplitude changes more or less rapidly, or in the opposite
direction, as a function of proper wavelength than in the standard
theory.

In conclusion, nonlinear dispersion alone is unlikely to have any
clearly discernable effects on the power spectrum of cosmological
perturbations. As suggested in Ref.~\cite{N00} and argued on the basis
of noncommmutative geometry in Ref.~\cite{CGS00}, a more promising
signature of Planck scale physics may be non-Gaussianity of the
fluctuation statistics.

\begin{acknowledgments}
J.C.N. would like to thank Achim Kempf for valuable discussions, and
acknowledges the support and hospitality of the DARC at Meudon and of
the University of Tours where this work was initiated in Dec. 2000.
R.P. is grateful to Amaury Mouchet for instructive discussions
concerning adiabaticity.
\end{acknowledgments}


\end{document}